\documentclass[useAMS,usenatbib]{mn2e}

\usepackage{psfig,epsfig,supertabular,wrapfig,placeins}
\usepackage{graphicx,amsmath,amssymb,subfigure,multirow,units}
\usepackage{marvosym,breqn}
\usepackage{graphics}
\usepackage{amssymb}
\usepackage{amsmath}
\usepackage{euscript}
\usepackage{setspace}
\usepackage{fancyhdr}
\usepackage{afterpage,rotating,url}
\usepackage{times}

\newcommand{\hatlas}{{\it H}-ATLAS}
\newcommand{\iras}{{\it IRAS}}

\newcommand{\gtsim}{\mbox{{\raisebox{-0.4ex}{$\stackrel{>}{{\scriptstyle\sim}}$}
}}}

\newcommand{\teff}{\ensuremath{T_{\mathrm{eff}}}}

\newcommand{\ldust}{\ensuremath{L_{\mathrm{dust}}}}
\newcommand{\lradio}{\ensuremath{L_{\mathrm{1.4\,GHz}}}}

\newcommand{\magphys}{{\sc MagPhys}}
 
\title[The temperature-dependent FIRC in {\it H}-ATLAS]{The
  temperature dependence of the far-infrared--radio correlation in the
  {\it Herschel}-ATLAS\thanks{{\it Herschel} is an ESA space
    observatory with science instruments provided by European-led
    Principal Investigator consortia and with important participation
    from NASA}}

\author[D.\,J.\,B.~Smith et al.]
       {D.\,J.\,B. Smith$^{1}$\thanks{E-mail:daniel.j.b.smith@gmail.com(DS)}, M.\,J. Jarvis$^{2,3}$, M.\,J. Hardcastle$^1$, M. Vaccari$^3$, N. Bourne$^4$, \newauthor L. Dunne$^5$, E. Ibar$^6$, N. Maddox$^7$, M. Prescott$^3$, C.~Vlahakis$^{8}$, S. Eales$^{9}$, \newauthor  S.\,J. Maddox$^5$, M.\,W.\,L. Smith$^{9}$, E.Valiante$^{9}$,  G.~de Zotti$^{10,11}$ \vspace{0.2cm}\\
    $^{1}$Centre for Astrophysics, Science \& Technology Research Institute, University of Hertfordshire, Hatfield, Herts, AL10 9AB\\
    $^{2}$Department of Astrophysics, Denys Wilkinson Building, Keble Road, Oxford, OX1 3RH \\
    $^{3}$Physics Department, University of the Western Cape, Private Bag X17, Bellville 7535, South Africa \\
    $^4$Institute for Astronomy, University of Edinburgh, Blackford Hill, Edinburgh EH9 3HJ\\
    $^5$Department of Physics and Astronomy, University of Canterbury, Private Bag 4800, Christchurch, 8140, New Zealand\\
    $^6$Instituto de F\'isica y Astronom\'ia, Universidad de Valpara\'iso, Avda. Gran Breta\~na 1111, Valpara\'iso, Chile\\
    $^7$Department of Astronomy, University of Cape Town, Private Bag X3, Rondebosch 7701, South Africa\\
    $^{8}$Joint ALMA Observatory\slash European Southern Observatory, Alonso de Cordova 3107, Vitacura, Santiago, Chile \\
    $^{9}$School of Physics and Astronomy, Cardiff University, Queen's Buildings, The Parade, Cardiff, CF24 3AA \\
    $^{10}$INAF-Osservatorio Astronomico di Padova, Vicolo Osservatorio 5, I-35122 Padova, Italy\\
    $^{11}$SISSA, Via Bonomea 265, I-34136 Trieste, Italy
}
\begin{document}

\date{\today}

\pagerange{\pageref{firstpage}--\pageref{lastpage}} \pubyear{2013}

\maketitle

\label{firstpage}

\begin{abstract}
We use 10,387 galaxies from the {\it Herschel} Astrophysical TeraHertz
Large Area Survey (\hatlas) to probe the far-infrared radio
correlation (FIRC) of star forming galaxies as a function of redshift,
wavelength, and effective dust temperature. All of the sources in our
250\,$\mu$m-selected sample have spectroscopic redshifts, as well as
1.4\,GHz flux density estimates measured from the Faint Images of the
Radio Sky at Twenty centimetres (FIRST) survey. This enables us to
study not only individual sources, but also the average properties of
the 250\,$\mu$m selected population using median stacking
techniques. We find that individual sources detected at $\ge 5\sigma$
in both the \hatlas\ and FIRST data have logarithmic flux ratios
(i.e. FIRC $q_\lambda$ parameters) consistent with previous studies of
the FIRC. In contrast, the stacked values show larger $q_\lambda$,
suggesting excess far-IR flux density\slash luminosity in 250\,$\mu$m
selected sources above what has been seen in previous analyses. In
addition, we find evidence that 250\,$\mu$m sources with warm dust
SEDs have a larger 1.4\,GHz luminosity than the cooler sources in our
sample. Though we find no evidence for redshift evolution of the
monochromatic FIRC, our analysis reveals significant temperature
dependence. Whilst the FIRC is reasonably constant with temperature at
100\,$\mu$m, we find increasing inverse correlation with temperature
as we probe longer PACS and SPIRE wavelengths. These results may have
important implications for the use of monochromatic dust luminosity as
a star formation rate indicator in star-forming galaxies, and in the
future, for using radio data to determine galaxy star formation rates.
\end{abstract}


\section{Introduction}
\label{sec:intro}

Until recently, the most widely used samples of galaxies selected at
far-infrared wavelengths in the local Universe have been derived from
wide-field observations using the {\it Infra-Red Astronomical
  Satellite} \citep[{\it IRAS};][]{neugebauer84}. Since the
\iras\ catalogues \citep{helou88,moshir92,wang09} are selected at
60\,$\mu$m, sampling the Wien region of the far-infrared spectral
energy distribution (SED), \iras-derived samples typically comprise
galaxies whose far-IR energy output is dominated by warm dust at
temperatures of $30-60$\,K. The vast increase in far-infrared
sensitivity at longer wavelengths afforded by the {\it Herschel Space
  Observatory} \citep[hereafter {\it Herschel;}][]{pilbratt10} allied
with its wide-field capabilities, have made it possible for the first
time to select large samples of local galaxies at wavelengths $\lambda
\gtsim 100\,\mu$m, where cool dust ($T \sim10-30$\,K) dominates the
SED \citep[e.g.][]{eales10,dunne11}.

The traditional model of the warm dust content of galaxies is that the
far-IR emission is associated with the stellar birth clouds, with O-
and B-type stars thought to dominate the dust heating \citep[e.g.][]{dejong84,helou85,sauvage92}. Since the stars
thought to be heating the dust are short-lived ($\sim 10-100$\,Myr),
the thermal far-infared emission has been widely used as a star
formation rate indicator \citep{kennicutt98,calzetti10,kennicutt12},
with the integrated dust emission frequently estimated from
sparsely-sampled photometry by assuming or deriving a model dust
spectrum, as the observations permit.

In the {\it Herschel} era, our understanding of the dust properties of
galaxies is being transformed, partially through enabling us to select
and study large samples of galaxies at longer far-IR wavelengths
\citep[e.g.][]{dunne11,smith12}. In particular, the unprecedented
sensitivity of {\it Herschel}-SPIRE at 250\,$\mu$m has highlighted the
prevalence of cooler dust in normal star-forming galaxies
\citep[though this was by no means a new idea; see
  e.g.][]{dejong84,kennicutt98,dunne00,dunne01,vlahakis05,draine07}. This
cool dust is thought to be associated with the diffuse interstellar
medium (ISM), and to be heated by older stars with longer lifetimes
than those which dominate the stellar birth clouds
\citep[e.g.][]{dale12,smith12,bendo12}; consequently, little
correlation between the cold dust luminosity and recent star formation
might be expected\footnote{Though in some cases stellar birth clouds
  may not be totally optically thick, and can ``leak'' UV photons;
  there may still be some correlation between heating of the diffuse
  ISM and recent star formation \citep[see e.g. ][]{popescu02}.}.

Though the most luminous radio sources are associated with non-thermal
synchrotron emission from active galactic nuclei (hereafter AGN), at
lower luminosities the source counts become dominated by star forming
galaxies \citep[e.g.][]{windhorst03,wilman08,massardi10,wilman10}. In
these sources, the relativistic electrons emitting the synchrotron
radiation are thought to have been accelerated by shocks resulting
from frequent supernovae, the end points of the same stars that
inhabit the stellar birth clouds and heat the dust (energy which is
then re-radiated in the far-infrared). These relativistic electrons
are thought to persist in the magnetic field of a galaxy for a few
tens of Myr emitting synchrotron radiation
\citep[e.g.][]{condon92}. As a result, radio luminosity has also been
frequently used as a star formation rate indicator
\citep{condon92,bell03}.

It is therefore not surprising that the far-infrared emission of
galaxies should show some relationship to the radio emission. What is
surprising about this relationship -- the far-infrared--radio
correlation, hereafter FIRC -- is that it is linear, that it shows
remarkably little scatter, and that it persists both over several
orders of magnitude in luminosity and for galaxies across the Hubble
sequence
\citep{vanderkruit71,dejong85,helou85,yunreddycondon01,garrett02}.

Several recent studies of the FIRC have investigated whether it
evolves with redshift, using data from the {\it Spitzer Space
  Telescope} and comparatively small areal coverage at radio
wavelengths in well-studied extra-galactic survey fields
\citep[e.g.][]{appleton04,frayer06,ibar08,murphy09,seymour09,michalowski10,sargent10,bourne11}
or for handfuls of sources using {\it BLAST} \citep{ivison10a}, {\it
  SCUBA} \citep{vlahakis07}, and {\it Herschel}
\citep{ivison10b}. These studies show apparently contradictory
results; \citet{seymour09} for example, found evidence for redshift
evolution, whereas \citet{appleton04} did not. Though the
observational evidence for an evolving FIRC is uncertain, it is
possible that the apparently contradictory results discussed above
might be reconciled with a non-evolving intrinsic FIRC, depending on
the effects of different selection functions in the studies mentioned
above (the aforementioned studies by Appleton et al. and Seymour et
al. exemplify this, being selected at far-IR and radio wavelengths,
respectively). For example evolution\slash uncertainty in the far-IR
SEDs of high-redshift sources might lead to uncertain $k$-corrections
\citep[e.g.][]{bourne11,seymour09}, and there is also the possibility
of contamination by low-luminosity AGN \citep[e.g.][]{verma02},
especially for the higher-redshift studies
\citep[e.g.][]{sargent10,jarvis10}.

There are many possible scenarios in which one might expect to observe
an evolving FIRC \citep[see ][for details]{lacki10a,lacki10b}. One
reason might be variations in the typical magnetic field strength in
galaxies (which would influence the radio emission but is unlikely to
affect the thermal dust emission). Another possible source of FIRC
evolution could be changing dust temperatures (e.g. due to the strong
temperature dependence of luminosity for modified black body
radiation, coupled with $k$-correction effects), or changes in the
dust distribution within a galaxy.

In what follows we build on the results of \citet{jarvis10} to revisit
the empirical properties of the FIRC at low redshift, taking
particular interest in the possible influence of the effective dust
temperature. We do this by taking advantage of the order of magnitude
increase in the area covered by the latest release of the {\it
  Herschel} Astrophysical TeraHertz Large Area Survey (hereafter
\hatlas), and the presence of shorter wavelength \hatlas\ data at 100
and 160\,$\mu$m, which were unavailable at the time of the previous
study. The presence of these data is particularly crucial for our
investigation, since a recent study by \citet{smith13} highlighted
their importance for determining isothermal dust temperatures for
local galaxies.

In section \ref{sec:observations} we describe the data used and our
sample selection, while in section \ref{sec:method} we describe our
methods of calculating the radio and far-infrared effective
temperatures\slash luminosities along with the FIRC. We present our
results in section \ref{sec:results} and make some concluding remarks
in section \ref{sec:conclusions}. We assume a standard $\Lambda$CDM
cosmology with $H_0 = 71$\,km\,s$^{-1}$\,Mpc$^{-1}$, $\Omega_M = 0.27$
and $\Omega_\Lambda = 0.73$ throughout.

\section{Observations}
\label{sec:observations}

\subsection{{\it Herschel}-ATLAS data}
\label{sec:hatlasdata}
This study is based on observations made with the {\it Herschel Space
  Observatory} as part of the {\it Herschel}-ATLAS survey
\citep[][Valiante et al., {\it in prep}]{eales10}. The
\hatlas\ catalogue consists of broad-band photometric imaging at 100
and 160\,$\mu$m from the PACS instrument \citep{poglitsch10}, and at
250, 350 and 500\,$\mu$m from the SPIRE instrument \citep{griffin10},
covering $\sim$161\,deg$^2$ over the three equatorial fields from the
Galaxy And Mass Assembly (GAMA) survey \citep{driver11}; further
details of the GAMA survey selection and strategy can be found in
\citet{baldry10} and \citet{robotham10}. Details of the
\hatlas\ map-making, source extraction and catalogue generation can be
found in \citet{ibar10,pascale11,rigby11} and Maddox et al., {\it in
  prep}. For this analysis, we use the far-IR flux densities in each
of the PACS\slash SPIRE bands taken from the current
\hatlas\ catalogue; the $5\sigma$ point source limits for each band in
order of increasing wavelength are 130, 130, 30.4, 36.9 \&\ 40.1\,mJy,
including confusion noise, with beam size between 9 and 35 arcsec
FWHM.

The current \hatlas\ catalogue recommends including calibration
uncertainties of 10 per cent of the measured flux density for the PACS
bands, and 7 per cent for the SPIRE bands, which we add in quadrature
to the estimated errors on the catalogue photometry. The
\hatlas\ catalogue has been cross-identified with $r$-band sources in
the Sloan Digital Sky Survey \citep[SDSS; ][]{york00}, using the
likelihood ratio technique discussed in \citet[][see also Bourne et
  al. {\it in prep}]{smith11}.

In this study, we include $>5\sigma$ 250\,$\mu$m sources with reliable
($R > 0.80$) optical counterparts, and robust spectroscopic redshifts
from GAMA in the latest \hatlas\ catalogue, which contains 13,084
sources which meet these criteria at $z < 0.5$ (though see section
\ref{sec:radiodata} below for a discussion of resolved sources).

\subsection{Radio data}
\label{sec:radiodata}

We use radio observations from the June 2013 release of the Faint
Images of the Radio Sky at Twenty-cm survey \citep[hereafter FIRST;
][]{becker95}, which covers 10,000\,deg$^2$ of the Northern Sky at
1.4\,GHz, with a typical RMS sensitivity of 0.15\,mJy\,beam$^{-1}$ in
the fields overlapping \hatlas. The \hatlas\ fields are also covered
by the NRAO VLA Sky Survey \citep[hereafter NVSS; ][]{condon98} with
45\,arc sec resolution and sensitivity of
0.45\,mJy\,beam$^{\mathrm{-1}}$, and there is also coverage of roughly
half of the \hatlas\ area at 325\,MHz taken using the Giant Metre-wave
Radio Telescope \citep{mauch13}, though the sensitivity varies
considerably.

In what follows we intend not only to probe the FIRC for individual
galaxies that are well detected at 1.4\,GHz, but also to statistically
probe the FIRC for the 250\,$\mu$m-selected population as a
whole. Therefore we do not use existing flux-density limited
catalogues but instead follow \citet{jarvis10}, and use the FIRST and
NVSS imaging data directly, producing cutout images 5\,arcmin on a
side centred on the optical positions associated to our 250\,$\mu$m
sources.

We used the cutout images to perform aperture photometry on the FIRST
maps, using 5\,arc sec radius circular apertures centred on the
positions of the reliable SDSS counterparts to the 250\,$\mu$m
sources. We derived uncertainties on each measurement by reading off
the value from the FIRST RMS maps, downloaded from the project
website\footnote{\url{http://sundog.stsci.edu/}}, and accounting for
the size of the aperture (i.e. converting from Jy beam$^{-1}$ to Jy
aperture$^{-1}$). To check that this rescaling was correct, we
systematically offset each aperture by 1 arcminute in a random
direction, and made a histogram of the resulting extracted aperture
fluxes, which were found to be consistent with the expected Gaussian
distribution. The FIRST flux densities derived in this way gave
excellent agreement (i.e. residuals consistent with the expected
distribution once the uncertainties are taken in to account) with the
values for the detected sources included in the June 2013 version of
the FIRST catalogue.

The potential advantage of the low resolution of NVSS relative to
FIRST \citep[i.e. possible greater resistance to resolved radio
  sources in our analysis;][]{jarvis10}, is offset by the lower
sensitivity (i.e. lower signal-to-noise in the stacks, and fewer
individual source detections), and by the fact that the NVSS images
are quantized in increments of 0.1\,mJy (which is comparable to the
average 1.4\,GHz flux density of an \hatlas\ 250\,$\mu$m source; see
section \ref{subsec:tdep}). In addition, reproducing the NVSS
catalogue fluxes for even the unresolved sources in our sample
requires corrections for fitting, confusion and {\it additive} clean
biases \citep{condon98}, with the latter bias being particularly
difficult to apply to stacked flux densities.

We also conducted a comparison between our FIRST aperture flux
densities and the NVSS catalogue values for the 78 sources with have
$\ge 5\sigma$ detections in each of the 250\,$\mu$m catalogue, our
FIRST aperture flux densities, and the NVSS catalogue (i.e. 250um
sources with NVSS catalogue flux densities $> 2.1\,$mJy). The
comparison reveals that the two sets of values are consistent
(i.e. the residuals are again consistent with the expected
distribution given the uncertainties).\footnote{We note that a subset
  of sixteen sources have a significant NVSS flux excess; upon visual
  inspection of the SDSS, FIRST and NVSS images for these sources, it
  becomes clear that in nine cases the NVSS excess is a result of
  blending with an unrelated nearby source in the large NVSS beam,
  while six are clearly double-lobed structures, and are recognised as
  AGN using our $q_{250} < 1.2$ criterion discussed in section
  \ref{sec:results}. The remaining source with an NVSS flux excess is
  also flagged as an AGN by our $q_{250}$ criterion, though unlike the
  previous six sources, this is not evident from its radio
  morphology. This offers further encouragement for our implementation
  of the \citet{hardcastle13} method of identifying AGN.} This
indicates that interferometer resolution effects (e.g. missing large
scale diffuse emission in the FIRST maps) are not an issue for our
sample. With these concerns in mind, we use our own FIRST aperture
flux densities directly measured from the cutout images in what
follows.

\section{Method}
\label{sec:method}

\subsection{Far-IR SED fitting}

\subsubsection{Isothermal SED fits}
\label{sec:isofit}

Our sample consists of sources with $> 5\sigma$ 250\,$\mu$m detections
(including confusion noise) with reliable optical counterparts and
spectroscopic redshifts. In order to derive the simplest possible
temperature estimates for our sample, we assume a single-component
modified blackbody model of the standard form:

\begin{equation}
  f_\nu \propto \frac{\nu^{3+\beta}}{\exp\left(\frac{h\nu}{kT}\right) - 1},
  \label{eq:gb}
\end{equation}

\noindent where $h$ is the Planck constant, $k$ is the Boltzmann
constant, and $T$ represents the dust temperature. The additional term
$\beta$ (known as the emissivity index) modifies the traditional
Planck function by assuming that the dust emissivity varies as a power
law of frequency, $\nu^\beta$. Following \citet{smith13}, we assume a
fixed $\beta$ of 1.82 \citep[similar to the value derived by
  e.g.][]{planck11}, and compare the five bands of \hatlas\ photometry
for each source with a grid of isothermal models with temperatures
between 5 and 60\,K, accounting for the transmission through the {\it
  Herschel} response curves. The temperature of the isothermal model
can be a useful empirical measure of the effective temperature of a
galaxy's dust SED \citep{smith13}.

\begin{figure}
  \centering
  \includegraphics[width=0.92\columnwidth]{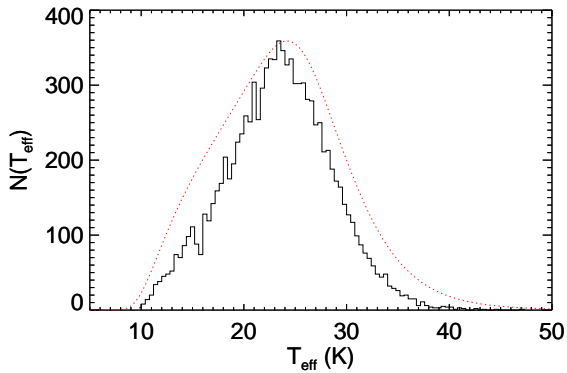}
  \caption{The effective temperatures recovered for our sample based
    on the isothermal SED fitting; median likelihood values are shown
    as the histogram (in black) and are overlaid with the summed
    temperature PDF for our sample (red dotted line) renormalised to
    have the same peak value as the histogram. The vast majority are
    in the range $10 < T_{\mathrm{eff}} < 40$\,K despite fitting with
    a temperature prior distribution that is flat between $5 < T <
    60$\,K. }
  \label{fig:thist}
\end{figure}

By recording the values of $\chi^2$ for every galaxy at every
temperature on the grid, we are able to build a marginalized
probability distribution function (hereafter PDF) for the effective
temperature of each object, \teff. We use our PDFs to generate median
likelihood estimates of \teff\ for each galaxy \citep[we use
  median-likelihood temperatures rather than best-fit estimates
  since][showed that they are less susceptible to bias when using
  \hatlas\ data to estimate them, though the difference is
  small\footnote{The small differences between the median-likelihood
    and best-fit estimates of \teff\ are shown in figure 14 of Smith
    et al. (2013), which contains simulations showing that the
    median-likelihood estimates have slightly narrower probability
    density contours than the best-fits (i.e. they are recovered more
    precisely), and that they exhibit better behaviour near the bounds
    of the temperature prior.}]{smith13}, with uncertainties derived
from the $16^{\mathrm{th}}$ and $84^{\mathrm{th}}$ percentiles of the
temperature PDF. To highlight the range of temperatures that we find
for galaxies in our sample, figure \ref{fig:thist} shows a histogram
of the median likelihood temperature estimates recovered for our
sample (in black), overlaid with the sum of each individual
temperature PDF (in red; which can give some indication of the
uncertainties on the median likelihood values).

\citet{smith13} used simulations designed to closely match the
\hatlas\ data set (including realistic models of both the instrumental
and confusion noise in each band) to determine how well effective
temperatures can be estimated for \hatlas\ sources. Using a superset
of the galaxies that we study here, \citet{smith13} found that
unbiased temperature estimates could be determined provided that PACS
data were included in the SED fitting, irrespective of whether a
particular source is formally detected by PACS or not (to
e.g. $3\sigma$). Since the \hatlas\ catalogue \citep[like the
  simulations in][]{smith13} contains flux density estimates in each
band irrespective of their formal significance, we use the catalogue
values as maximum likelihood estimates, and consequently we do not
have to use upper limits in our fitting. Using the results of
\citet{smith13} we find that the average uncertainty in \teff\ is
$\Delta \teff \slash \teff \approx 0.166$ across the full range of
temperatures recovered here (figure \ref{fig:thist}), ranging from
$\Delta \teff \sim 2$\,K at 10K to $\Delta \teff \approx 6$\,K at
40\,K.

\subsubsection{Estimating Integrated dust luminosities}
\label{sec:ldust_est}

Dust in galaxies is not isothermal; though the dust SED generally
peaks at $\lambda \gtsim 80$\,$\mu$m, there is also a varying
contribution to the integrated dust luminosity from very small, hot
grains that are bright in the mid-infrared \citep[e.g.][]{yang07}, and
this contribution can exceed 0.3\,dex in luminosity. Though the
\hatlas\ fields are covered by mid-infrared data from the Wide-field
Infrared Survey Explorer \citep[WISE;][]{wright10}, and extensive
efforts have been made to provide aperture-matched WISE photometry for
\hatlas\ sources \citep{cluver14}, the preliminary catalogues contain
detections in the W4 (22\,$\mu$m) band for $< 5$ per cent of our
sample. As a result, the vast majority of our integrated dust
luminosities are strongly model-dependent, limited in precision due to
the variable contribution of the mid-infrared to the integrated
luminosity, and subject to possible bias that is difficult to quantify
(see appendix \ref{sec:kcor} for more details).

For these reasons, we do not use integrated dust luminosities in our
analysis. However, our simplest isothermal estimates of
\ldust\ (integrated between 3-1000\,$\mu$m) show that the majority of
our sample have far-infrared luminosities in the star-forming galaxy
regime, with $10.0 < \log_{10} (\ldust \slash L_\odot ) < 11.0$,
though there is a substantial minority in the luminous infrared galaxy
category; $\log_{10} (\ldust \slash L_\odot ) > 11.0$ \citep[see
  also][]{smith13}.

\subsubsection{Estimating FIR monochromatic luminosities}
\label{subsubsec:mono_estimates}

Whilst the integrated dust luminosities for our sample show dependence
on the choice of SED template used, our monochromatic luminosity
estimates are much more robust. This is a result of the high-quality
\hatlas\ far-IR photometry (i.e. the data are uniform, and the
photometric scatter is small), and the generally precise
$k$-corrections ($k_\lambda$) that we are able to derive for our $z <
0.5$ sample. To determine $k_\lambda$ we require a best estimate of
the underlying dust SED; for this task we use the best-fit SED for
each galaxy from the \citet[][hereafter SK07]{sk07} model library. We
use SK07 SEDs because they are able to recover good fits to the
\hatlas\ data across the full range of effective temperatures shown in
figure \ref{fig:thist}, and because they are more realistic than the
isothermal models.

\begin{figure}
  \centering
  \includegraphics[width=0.92\columnwidth]{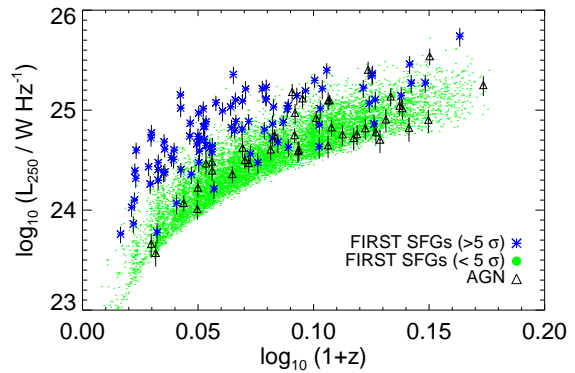}
  \caption{Relationship between redshift and 250\,$\mu$m monochromatic
    luminosity, $L_{250}$, for galaxies where the radio emission is
    dominated by star formation according to FIRST (blue asterisks),
    by AGN (black triangles) and the 250\,$\mu$m galaxy population
    that is not detected by FIRST (in green). The AGN have been
    identified using a method analogous to that of
    \citet{hardcastle13}; see text for more details.}
  \label{fig:z_ldust_l250}
\end{figure}

The accuracy of our $k_\lambda$ is discussed in appendix
\ref{sec:kcor}, but to summarise, the median uncertainty on the
individual $k$-corrections ranges from $\sim 14$\,per cent at
250\,$\mu$m to $\sim 25$\,per cent at 100\,$\mu$m, though the
uncertainty on $k_{100}$ at the coldest temperatures is rather
larger. In what follows we account for the uncertainty on $k_\lambda$
by adding the estimated errors in quadrature with the uncertainties on
the individual flux densities taken from the \hatlas\ catalogue. The
monochromatic $L_{250}$ for galaxies in our sample as a function of
redshift are shown in figure \ref{fig:z_ldust_l250}.

\subsection{Calculating 1.4 GHz luminosity}
\label{sec:radiolum} 

Since we use FIRST aperture flux densities in our analysis for the
reasons discussed in section \ref{sec:radiodata}, we minimize the
influence of resolved sources on our studies of the FIRC by excluding
all sources with SDSS major axes $> 10$\,arc sec. We $k$-correct the
1.4\,GHz luminosity density to the rest frame for each source by
assuming spectral indices which are randomly drawn from a Gaussian
distribution centred about $\alpha = 0.71$ with an RMS of 0.38
\citep[using the convention that $S_\nu \propto
  \nu^{-\alpha}$;][]{condon92}. These values for $\alpha$ were derived
by \citet{mauch13} using 90\,deg$^2$ of 325\,MHz and 1.4\,GHz data
within the \hatlas\ fields. Since the sources in this study are all at
$z < 0.5$, the derived $k$-corrections are small (the median
uncertainty on this $k$-correction for our sample is $\sim 6.3$ per
cent), however we account for this additional source of uncertainty in
two ways. Firstly, we repeated our analyses 100 times using random
draws from the Gaussian $\alpha$ distribution, finding that our
results are unchanged within the errors. Secondly, we propagate our
estimates of the uncertainty on each individual $k$-correction
(derived by determining the standard deviation of $k_{\mathrm{1.4 GHz}}$ as a
function of redshift) through to the derived luminosities by adding
them in quadrature with the uncertainties on the 1.4\,GHz flux density
estimates themselves.

\begin{figure}
  \centering
  \includegraphics[width=0.92\columnwidth]{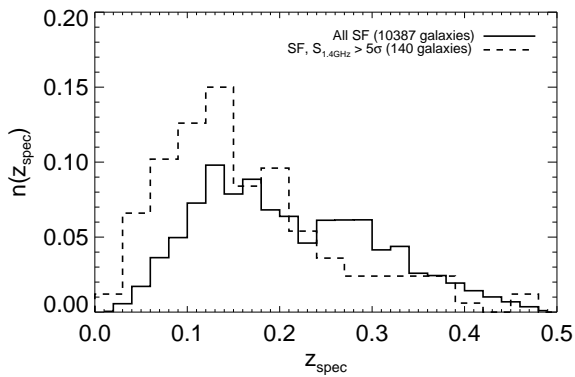}
  \caption{The redshift distribution of the 10,387 250\,$\mu$m
    selected \hatlas\ sample with spectroscopic redshifts $z_{spec} <
    0.5$ (solid lines) and isophotal major axes $< 10$\,arc sec. Also
    overlaid (dashed line) is the redshift distribution of the subset
    of 140 galaxies which have $> 5\sigma$ detections at 1.4\,GHz in
    FIRST; see section \ref{sec:radiolum} for details. }
  \label{fig:zdist}
\end{figure}

\begin{figure}
  \centering
  \includegraphics[width=0.92\columnwidth]{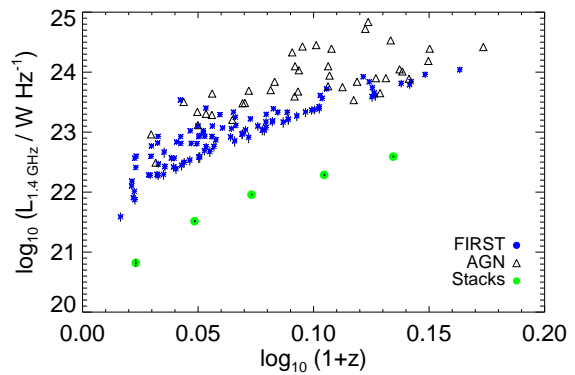}
  \caption{The 1.4\,GHz luminosity density of sources in our sample as
    a function of redshift, with those sources in the starburst regime
    detected by FIRST shown in blue. The median stacked 1.4\,GHz
    luminosity density for all 250\,$\mu$m sources in our sample,
    divided between five bins of redshift are shown as the green
    circles. Sources with a substantial AGN contribution to
    \lradio\ are shown as black triangles; see section
    \ref{sec:results} for details of their identification.}
  \label{fig:z_lradio}
\end{figure}

The spectroscopic redshift distributions of 250\,$\mu$m sources with
$\ge 5\sigma$ 1.4\,GHz detections, and all 250\,$\mu$m detected
sources in our sample are shown in figure \ref{fig:zdist}, with both
distributions peaking at $z_{\mathrm{spec}} \approx 0.13$. In figure
\ref{fig:z_lradio} we show \lradio\ as a function of redshift, with
the 250\,$\mu$m selected star-forming galaxies with SNR $> 5$ at
1.4\,GHz shown in blue, and the stacked values for our whole
250\,$\mu$m sample in bins of redshift shown in green. The stacked
values are obtained by simply taking the median of the individual
\lradio\ values (whether statistically significant or not) in bins at
intervals of 0.03 in $\log_{10} (1+z)$, except for the highest-$z$ bin
which contains all sources at $\log_{10} z_{spec} > 0.012$.

\subsection{Calculating the far-infrared--radio correlation}
\label{subsec:firc_calc}

The dimensionless parameter describing the FIRC, $q$, is defined as
the logarithmic ratio of the far-infrared luminosity \ldust,
integrated between 3-1000\,$\mu$m in the rest-frame, to the rest-frame
1.4\,GHz $k$-corrected luminosity density \lradio, such that:

\begin{equation}
  q = \log_{10}\left ( \frac{L_{\mathrm{dust}}/3.75 \times
    10^{12}}{L_{1.4\,GHz}} \right ).
  \label{eq:q}
\end{equation}

\noindent Here $3.75 \times 10^{12}$ is the frequency corresponding to
80\,$\mu$m, making $q$ dimensionless. This is equivalent to the
definitions given by \citet{helou85,bell03} and \citet{ivison10b}, who
quote the logarithmic ratio in terms of flux densities rather than
luminosities. In \hatlas\ our five bands of far-infrared observations
sample wavelengths near (and long-ward of) the peak of the dust SED for
local galaxies. Since, as discussed in section \ref{sec:ldust_est},
our integrated dust luminosity estimates are not currently accurate
enough to probe the small variations in $q$ found by
\citet{ivison10a}, here we focus on the monochromatic equivalents,
$q_{\lambda}$, such that:

\begin{equation}
  q_{\lambda} = \log_{10}\left (
  \frac{L_\lambda}{L_{\mathrm{1.4\,GHz}}} \right ),
  \label{eq:q250}
\end{equation}

\noindent where $\lambda$ can be any of the PACS\slash SPIRE
wavelengths, e.g. $q_{250}$. We calculate errors on $q_{\lambda}$ by
propagating the errors from the input flux densities and $k_\lambda$
(though the latter source of error is typically small except for the
$T < 20$\,K sources at 100\slash 160\,$\mu$m; see section
\ref{subsubsec:mono_estimates} and appendix \ref{sec:kcor} for
details).

Since only a small proportion (140\slash 10,387) of the sources in our
sample have $> 5\sigma$ detections at 1.4\,GHz, we also calculate
stacked $q_\lambda$ for all galaxies in our 250\,$\mu$m-selected
sample. Following \citet{bourne11}, we derive stacked $q_\lambda$ for
our 250\,$\mu$m-selected sample by calculating the median
$k$-corrected flux density in each PACS\slash SPIRE band, then
dividing by the median stacked $k$-corrected 1.4\,GHz flux density,
according to equation \ref{eq:q250}. We use median stacking since this
method is more resistant to the effects of outliers (due to
e.g. residual AGN contamination) in the individual flux density
estimates than using the mean. We note that we stack on the
$k$-corrected PACS\slash SPIRE\slash FIRST flux densities themselves
(i.e. the values which we derive in sections
\ref{subsubsec:mono_estimates} and \ref{sec:radiolum}) rather than by
producing stacked images (which are more difficult to correctly
$k$-correct). We determine the uncertainties associated with each
median stacked flux density by bootstrapping based on 1000 re-samples
of the values in each stack; our simulations show that this
non-parametric method gives excellent agreement with results obtained
using the median statistics method from \citet{gott01} used by
\citet{bourne11}, and that it accounts naturally for the uncertainties
on the individual values.

\subsection{Summary of sample selection}

Our study is based on 13,084 galaxies with a signal to noise ratio
$>5$ at 250\,$\mu$m, reliable ($R > 0.80$) counterparts in the
\citet{smith11} likelihood ratio analysis, and spectroscopic redshifts
$0.00 < z_{\mathrm{spec}} < 0.50$. After removing sources with
isophotal semi-major axes $> 10$\,arc sec, 11,389 sources remain, of
which 10,437 have good fits to our isothermal model (i.e. they have
reduced $\chi^2 < 2.0$). As we will discuss in section
\ref{subsec:agnremoval} below, we also remove an additional 50 sources
that we classify as AGN, leaving a sample of 10,387 galaxies on which
our results are based.

\section{Results}
\label{sec:results}

\subsection{The monochromatic FIRC and its redshift dependence}
\label{subsec:agnremoval}

Before searching for the presence of a correlation between the
far-infrared and radio emission of the galaxies in our sample, we must
first account for the potential presence of radio-loud AGN; these
sources are likely to have excess emission at 1.4\,GHz and may bias
our estimates of $q_\lambda$ to lower values. Among the radio sources
detected at $> 5\sigma$, we identify those likely to be dominated by
powerful AGN rather than star formation using a method similar to that
in \citet[][]{hardcastle13}.\footnote{See also
  \citet{condon91,sanders96,yunreddycondon01,verma02}.} This method --
which enables us to specify a threshold in $q_{250}$ to identify
AGN-dominated sources -- was originally based on classifying
FIRST-detected radio sources as star-forming or AGN using their
optical spectra, and comparing the classifications with $q_{250}$.

\begin{figure}
  \centering
  \includegraphics[width=0.92\columnwidth]{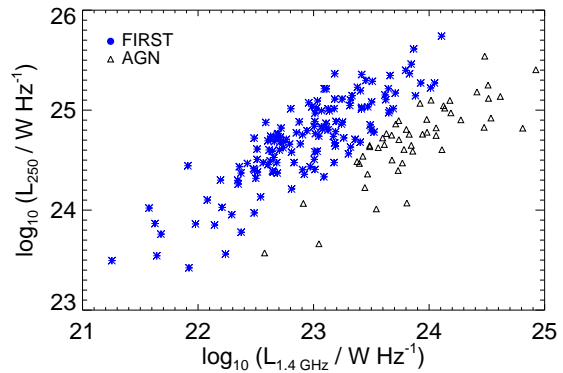}
  \caption{The relationship between \lradio\ and $L_{250}$ for the
    star-forming galaxies in our sample detected at $> 5\sigma$ at
    1.4\,GHz (in blue). AGN identified using a method based on that in
    \citet{hardcastle13} are indicated by the black triangles.}
  \label{fig:l1400_ldust_l250}
\end{figure}

We identify 50 sources which have $q_{250} < 1.2$ as AGN (black
triangles in figure \ref{fig:l1400_ldust_l250}), and disregard them
from what follows.\footnote{The potential downside of this approach is
  that we may miss outliers in the parameter of interest, however
  previous work on complete samples suggests that we expect very few
  of these, and our median stacking provides a degree of resistance to
  the impact of this effect.} In contrast to a more traditional
approach to dealing with AGN contamination, such as removing all
sources with the highest radio luminosities
\citep[e.g.][]{mauch07,jarvis10,lemaux13}, this method allows us to
keep the most luminous star-forming galaxies in our
250\,$\mu$m-selected sample, and removes obvious powerful AGN with
lower radio luminosities (indicated by the black triangles in figures
\ref{fig:z_ldust_l250}, \ref{fig:z_lradio} and
\ref{fig:l1400_ldust_l250}).

While this method has been shown to identify the most obvious powerful
AGN, less obvious low-luminosity AGN are harder to identify, such as
those that are not detected at $> 5\sigma$ in the FIRST data, or
inefficient AGN which may not have been apparent in the optical
spectra used in \citet{hardcastle13} when generating the $q_{250}$
criterion (though the latter tend to be ``red and dead'', and so they
are unlikely to meet our 250\,$\mu$m selection criterion). As a
result, though our 250\,$\mu$m-selection should ensure that our sample
is dominated by star-forming galaxies (and the fact that $< 1$ per
cent of our sample are flagged as obvious AGN reflects this) it is
possible -- or even likely -- that some low-luminosity AGN are
present. Though the influence of low-luminosity AGN is likely to be
strongest on the individual data points (due to our use of median
stacked flux densities for calculating $q_\lambda$, which should be
resistant to moderate levels of AGN contamination), it is still
possible that a fraction of the 250\,$\mu$m sources in our sample
harbour AGN which may enhance the average radio luminosity at a lower
level. The effect of such an AGN contribution would be to bias our
$q_\lambda$ estimate to lower values, and it is a potential effect
that we must bear in mind in what follows. Though removing the 50
sources with $q_{250} < 1.2$ from our stacks has negligible impact
upon the results, these represent a sizeable fraction of the $5\sigma$
FIRST detected sources ($\sim 32$\,per cent), higlighting the
importance of our stacking analysis.

In figure \ref{fig:l1400_ldust_l250} we show $L_{250}$ as a function
of \lradio\ for the 1.4\,GHz detected sources; we observe a strong
correlation between the radio and far-infrared, consistent with
previous studies
\citep[e.g.][]{garn09a,garn09b,bourne11,ivison10a,ivison10b,jarvis10},
and the sources that we identify as AGN are clearly offset to higher
\lradio, as we would expect.

\begin{figure}
  \centering
  \includegraphics[width=0.92\columnwidth, trim=0.8cm 0cm 0cm 0.90cm, clip=true]{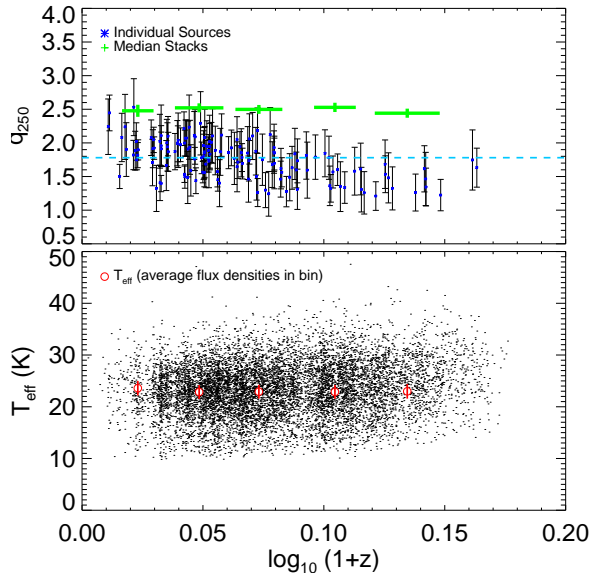}
  \caption{{\bf Top:} The redshift dependence of the 250\,$\mu$m
    monochromatic far-infrared--radio correlation. $> 5\,\sigma$ FIRST
    sources are shown in blue, with black error bars derived from
    propagating through the errors on the input luminosities
    ($L_{250}$ and \lradio) and accounting for the uncertainties on
    the $k$-corrections. The median stacked $q_{250}$ values for all
    250\,$\mu$m sources in bins of redshift are shown as the green
    error bars, with the vertical error bars derived by bootstrapping,
    re-sampling each bin 1000 times. The blue horizontal dashed line
    indicates the best-fit $q_{250}$ from \citet{jarvis10}. {\bf
      Bottom:} \teff\ versus spectroscopic redshift for the individual
    sources in our sample (black points) and for the median-stacked
    flux densities in each redshift bin (red circles with error
    bars). }
  \label{fig:q_z}
\end{figure}

In the top panel of figure \ref{fig:q_z}, we parameterise the sources
in figure \ref{fig:l1400_ldust_l250} using equation \ref{eq:q250}, and
show the 250\,$\mu$m far-infrared--radio correlation parameter
$q_{250}$ as a function of redshift, with the best estimate from
\citet{jarvis10} overlaid as the dashed horizontal light blue
line. Though the individual monochromatic $q_{250}$ for the $>5\sigma$
1.4\,GHz detected sources show good agreement with the best fit from
\citet{jarvis10}, our stacked estimates in bins of redshift (shown as
the green points with error bars) are offset to higher $q_{250}$. That
our stacked values are higher than those for the individual detections
is likely a result of the fact that we are determining $q_{250}$
statistically for the full 250\,$\mu$m galaxy population. The median
1.4\,GHz flux density of our sample is around 150\,$\mu$Jy, making the
average 250\,$\mu$m selected galaxy detectable only in the most
sensitive radio data. The median stacked $q_{250}$ that we observe
($q_{250} \approx 2.61$) is larger than the value found by
\citet{ivison10b} using a 250\,$\mu$m-selected sample of 22 BLAST
sources ($q_{250} = 2.18 \pm 0.28$), though the values are consistent
within $\sim1.5\sigma$.

The stacked $q_{250}$ values shown in the top panel of Figure
\ref{fig:q_z} reveal no evidence for evolution with redshift, with the
dearth of detected sources at high $q_{250}$ due to the relative lack
of sensitivity of FIRST compared to our {\it Herschel} data; we only
detect the brightest 1.4\,GHz sources in the FIRST data, and this is
reflected by the stacks being offset to higher $q_\lambda$ than the
1.4\,GHz detections. This result -- average stacked $q_\lambda$ being
larger than the individually detected data points -- highlights the
benefits of using a stacking analysis to probe the general population
rather than focusing solely on the brightest radio continuum sources
in the distribution; we shall return to this offset in the next
section.

As well as searching for evolution in $q_\lambda$ with redshift, in
the lower panel of figure \ref{fig:q_z} we search for variation in
temperature with redshift. Spearman's rank tests reveal no significant
evidence for correlation between temperature and redshift for the
individual galaxies (black points; the asymmetric error bars on
\teff\ are not shown for clarity). In addition, if we fit isothermal
models to the stacked $k$-corrected flux densities in bins of
redshift, we find that the best-fit temperature in each bin is
consistent with no evolution (red circles with error bars in the lower
panel of figure \ref{fig:q_z}).

\subsection{The temperature dependence of the FIRC in \hatlas}
\label{subsec:tdep}

\begin{figure}
  \includegraphics[width=0.95\columnwidth]{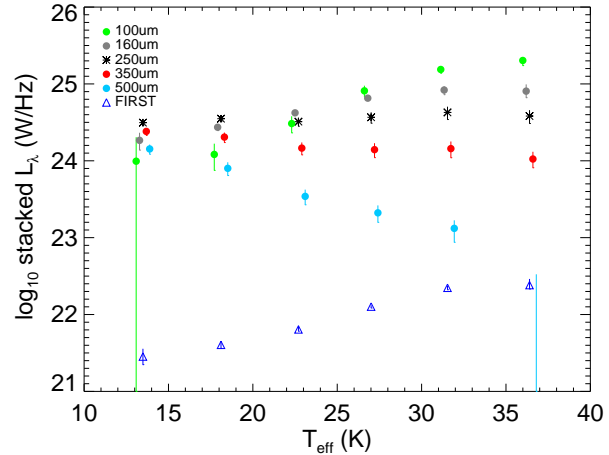}
  \caption{The median-stacked monochromatic luminosity density in the
    PACS\slash SPIRE bands (in green, grey, black, red and light blue,
    from 100-500\,$\mu$m, respectively), and the FIRST data at
    1.4\,GHz (in blue). The errors have been estimated based on
    bootstrapping 1000 resamples of the data, and added in quadrature
    with the uncertainties on the $k$-corrections as a function of
    temperature. The median-stacked monochromatic luminosity density
    at 500\,$\mu$m in the hottest bin is negative (though not
    statistically significant), and thus only the upper part of the
    error bar is visible on the logarithmic vertical axis. The
    individual PACS/SPIRE data points have been slightly offset at
    intervals of 0.2\,K in the abscissa to prevent the error bars from
    obscuring one another. There are 641, 1973, 3872, 2912, 843 and
    146 galaxies in each bin, with the numbers listed in order of
    increasing temperature. }
  \label{fig:fluxstacks_temp}
\end{figure}

Before analysing the $q_\lambda$, we first show the median-stacked
monochromatic luminosity densities in Figure
\ref{fig:fluxstacks_temp}, in bins of temperature. The PACS and SPIRE
values are shown with error bars indicating the sum in quadrature of
the bootstrapped errors on the median fluxes and the uncertainties on
$k_\lambda$, while the 1.4\,GHz values (blue triangles) include only
bootstrapped errors (though as we discussed in section
\ref{sec:radiolum}, 100 Monte-Carlo realisations of the 1.4\,GHz
$k$-corrections revealed that this source of uncertainty is smaller
than the bootstrapped errors on the stacked values).

Interestingly, we find that the stacked $k$-corrected 1.4\,GHz
luminosity densities (blue triangles in figure
\ref{fig:fluxstacks_temp}) reveal a clear increase with the effective
temperature of the dust SED (the increase is around an order of
magnitude between the coldest and warmest bins of our sample). This is
indicative of warmer galaxies hosting larger levels of star formation,
as expected based on the FIRC combined with the far-infrared
luminosity--temperature (``L--T'') relation found by several previous
studies \citep[e.g.][]{chapman03,hwang10,smith13}. We observe this
relationship at radio wavelengths for the first time. The lack of
temperature evolution in our sample (highlighted in the lower panel of
figure \ref{fig:q_z}) suggests that this trend, the
radio-luminosity--temperature relation, is not simply the result of
redshift\slash luminosity effects.

\begin{figure*}
  \includegraphics[width=2.0\columnwidth, trim=0cm 0cm 0cm 0.8cm,
    clip=true]{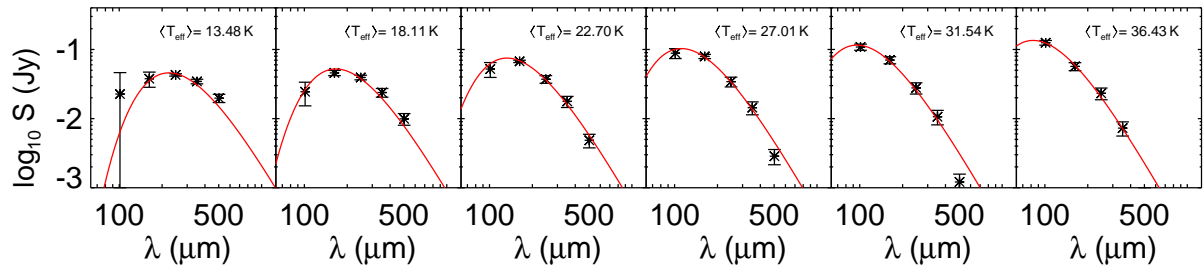}
  \caption{The $k$-corrected stacked flux densities (black asterisks
    with error bars) in each of the six temperature bins, overlaid
    with an isothermal dust SED (in red). The median temperature of
    the individual galaxies in each bin is used to define the
    isothermal model, and is displayed in the top right corner. }
  \label{fig:fluxes_tbins}
\end{figure*}

Figure \ref{fig:fluxes_tbins} shows the $k$-corrected stacked flux
densities (in black) for the same six temperature bins shown in figure
\ref{fig:fluxstacks_temp}, overlaid (in red) with isothermal models
corresponding to the median temperature of the galaxies in that
bin. These panels highlight how well the stacked flux densities in
figure \ref{fig:fluxes_tbins} are described by the isothermal model
that we use to estimate temperatures.

\begin{figure}
  \centering
  \includegraphics[width=0.95\columnwidth]{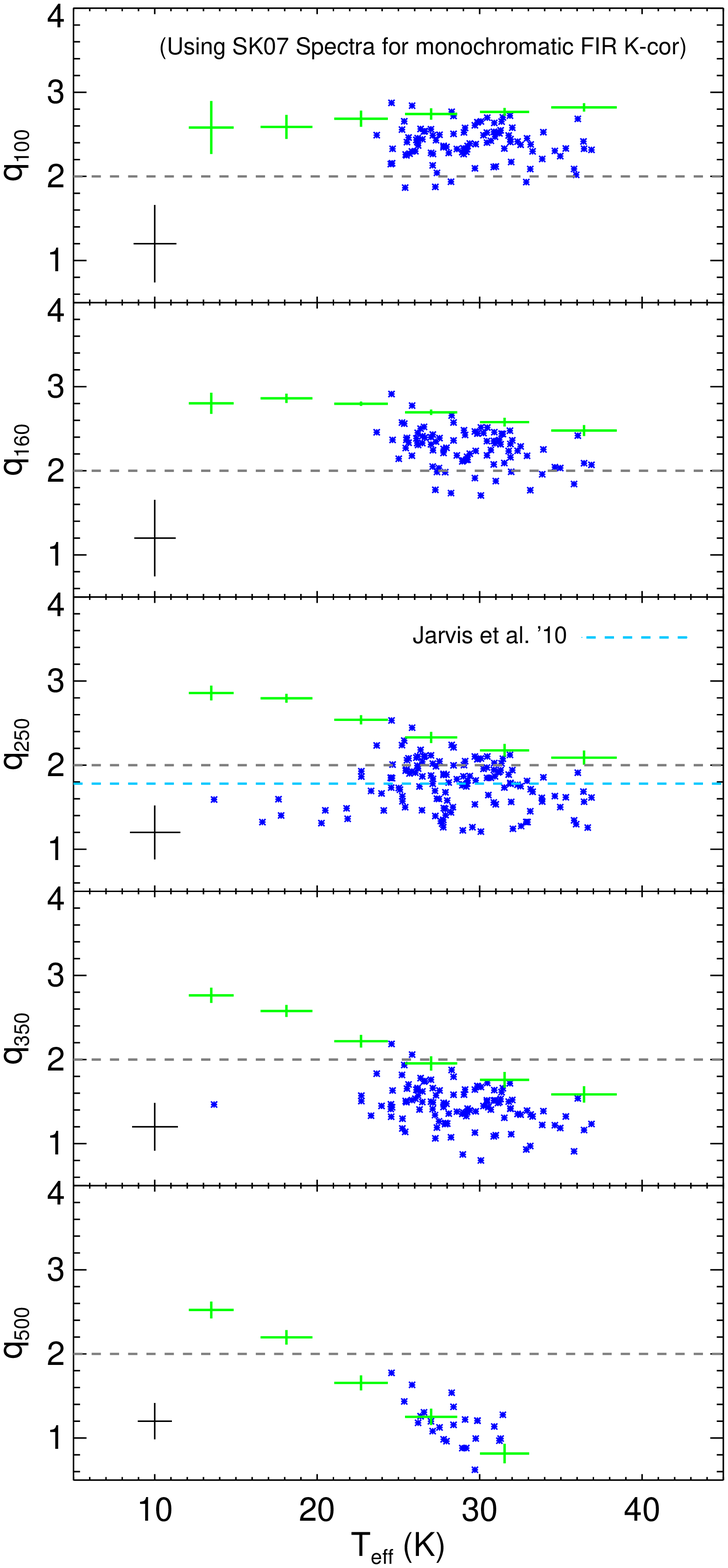}
  \caption{Variation of the monochromatic FIRCs as a function of
    effective temperature. In addition to the initial catalogue
    selection (i.e. $\ge 5\sigma$ at 250\,$\mu$m, $\ge 5\sigma$ at
    1.4\,GHz), we require a minimum $5\,\sigma$ detection in the
    relevant band for the individual monochromatic $q_\lambda$ to be
    plotted; this is reflected in the lower number of blue points in
    the $q_{500}$ panel compared with the other bands. The stacked
    $q_\lambda$ are shown as the green error bars, while the
    horizontal grey dashed lines indicate $q_\lambda = 2$, to guide
    the eye. The stacked $q_{500}$ in the warmest temperature bin is
    negative (as discussed in the caption to figure
    \ref{fig:fluxstacks_temp}) and so not displayed here. The median
    error uncertainty for the individually-detected sources is shown
    by the black error bar in the lower left corner of each panel. }
  \label{fig:t50_q_mono}
\end{figure}

In figure \ref{fig:t50_q_mono} we show the variation of the
monochromatic FIRC parameters as a function of temperature ($q_{100}$,
$q_{160}$, $q_{250}$, $q_{350}$, $q_{500}$ from top to bottom,
respectively). Whilst in every panel we require a $\ge5\sigma$
detection at 250\,$\mu$m, we also require a $\ge 5\sigma$ detection in
the particular band shown for an individual source to be included. We
display the sources detected at $\ge 5\sigma$ in FIRST as the blue
crosses, and overlay the stacked $q_\lambda$ in bins of temperature as
green error bars, derived as discussed in section
\ref{subsec:firc_calc}, and based on the median stacked luminosity
densities shown in figure \ref{fig:fluxstacks_temp}. As in figure
\ref{fig:q_z}, we find that if we consider only the 250\,$\mu$m
sources which are detected at 1.4\,GHz, we recover values of
$q_\lambda$ offset from the values stacked across the whole
250\,$\mu$m selected population (shown in green in figure
\ref{fig:t50_q_mono}).

At the PACS wavelengths, the stacked values reveal that $q_{100}$
shows little evidence for variation as a function of effective dust
temperature, however as we move to progressively longer wavelengths
$q_{160}$, $q_{250}$, $q_{350}$ and $q_{500}$ show increasingly strong
negative correlations with the effective temperature of the far-IR
SED. A previous study of the temperature dependence of the FIRC
\citep{ivison10a} found evidence for weak negative correlation between
the integrated $q$ and dust temperature, concentrating on a
250\,$\mu$m-selected sample of 128 galaxies with a mean redshift $\sim
1$. The \citet{ivison10a} 250\,$\mu$m selection at $z\sim 1$
corresponds to selection at $\sim 100-160$\,$\mu$m (near the peak of
the dust SED) in the rest frame, where our monochromatic values show
broadly consistent behaviour. We are currently unable to probe
integrated $q$ using the \hatlas\ data set, though the fact that $q$
in \citet{ivison10a} and $q_{100}$\slash$q_{160}$ in this study are
approximately temperature independent offers some encouragement for
the use of \ldust\ as an SFR indicator in certain situations
\citep[though see also][]{dacunha12,smith12,hayward14,rowlands14}.

We must also consider the possible influence of contamination in the
250\,$\mu$m cross-identifications upon our results. \citet{bourne14}
suggest that the 250\,$\mu$m-SDSS cross-identification analysis in
\citet{smith11} may have overestimated the number of reliable
associations for the coldest sources in \hatlas\ with $T <
15$\,K. However, the trends that we see in figure \ref{fig:t50_q_mono}
all remain unchanged even in the implausible extreme scenario in which
erroneous cross-IDs dominate the sources in the coldest bin, therefore
our results and conclusions are robust to these effects.

\subsection{Discussion}

If we assume a simple two-component model of the dust SED, similar to
that proposed by e.g. \citet{charlot00} and implemented in
\magphys\ \citep{dacunha08}, we expect the 100\,$\mu$m PACS
observations to sample emission associated with the warm stellar birth
cloud (BC) dust component of a galaxy, and to show a strong
correlation with the most recent burst of star formation due to
heating dominated by short-lived OB stars. In contrast, we expect that
the longer wavelength observations should be dominated by emission
from colder dust in the ambient interstellar medium (ISM), perhaps
heated by more evolved stars and more weakly correlated with recent
star formation. The influence of the ISM on the FIRC is highlighted by
\citet{vlahakis07}, who attributed the larger variation in the FIRC at
850\,$\mu$m compared to 60\,$\mu$m to the influence of a varying cold
dust component. In the ISM$+$BC scenario, it is relatively
straightforward for a comparatively low-mass component of warm dust in
the BC to become more luminous than the cooler, more massive, ISM
component at 250\,$\mu$m due to the strong temperature dependence of
[modified] black body radiation. Furthermore, black body physics
implies that the shorter wavelengths (i.e. the BC) fade more rapidly
after the truncation of star formation than the longer wavelengths as
the gas cools. The exact temperature dependence of the monochromatic
$q_\lambda$ may also be affected by differing decay times for the dust
SED and the synchrotron energy density following the truncation of
star formation, a possibility investigated in detail by
\citet{lacki10a} and \citet{lacki10b}, to which we refer the reader
for further details.

In this simple two-component scenario galaxies can meet our
250\,$\mu$m selection criterion by having an SED dominated by a
low-mass, warm BC component, through having a dominant high-mass cold
ISM component, or some mix of the two as we move from the hottest to
the coldest SEDs in our sample. An estimate of the proportion of the
total dust luminosity contributed by the ISM -- ``$f_\mu$'' -- is
produced by the {\sc magphys} SED fitting for \hatlas\ galaxies in
\citet{smith12}. Though the individual $f_\mu$ estimates have large
uncertainties, we determine the median $f_\mu$ in temperature bins,
and find weak evidence for decreasing $f_\mu$ with increasing
effective temperature. This suggests that the BC are more dominant at
warmer temperatures (as we would expect), but the weak trend hints
that the true situation is likely more complicated than the simple two
component model, and that the temperature information is inadequate to
determine the relative mix of BC and ISM emission in galaxies on its
own.

The physical difference between the two extremes in temperature for
this simple model, and the 250\,$\mu$m selection, may be the dominant
forces in the variation of $q_{500}$ (since the 500\,$\mu$m luminosity
varies much more than the luminosity at 1.4\,GHz across our
temperature range). In contrast, the 100\,$\mu$m data are generally
dominated by dust heated in the most recent burst of star formation,
due to the much stronger dependence of 100\,$\mu$m luminosity on
temperature than on dust mass, and so less variation in $q_{100}$
might be expected if we assume that the 1.4\,GHz emission is also
related to the most recent burst. This highlights that studies of the
FIRC -- or investigations assuming a constant FIRC -- must be wary of
temperature effects if wavelengths away from the peak of the dust SED
are used.

Though we account for the presence of obvious AGN contamination using
the method of \citet{hardcastle13}, it is possible that our sample
contains residual low-level AGN contamination, which has the potential
to bias our results to lower $q_\lambda$. In an attempt to simulate
the influence of residual AGN-contamination on our results, we perform
a simple test; we arbitrarily assume that the 1.4\,GHz flux density of
a random 15 per cent of our sample is dominated by AGN. We ``correct''
these values by artificially replacing them with values drawn from a
random distribution with a median of zero and standard deviation equal
to the local RMS flux density from the FIRST maps appropriate for each
source (i.e. with values consistent with zero). Repeating our analysis
using these artifical ``AGN-subtracted'' values results in our FIRC
estimates shown in figure \ref{fig:q_z} and \ref{fig:t50_q_mono} being
offset to larger $q_{\lambda}$ by $\sim 0.1$\,dex.\footnote{Performing
  the same simulation, but instead assuming the extreme case in which
  50 per cent of our sample has 1.4\,GHz flux density entirely due to
  AGN, and replacing those values in the same way (i.e. making them
  consistent with zero), alters our results in figures \ref{fig:q_z}
  and \ref{fig:t50_q_mono} to larger $q_\lambda$ by $\sim 0.3$\,dex.}
We conclude that though it is possible that residual AGN exert some
influence on the results of our stacking analyses, this influence is
likely to be small.

Our results highlight potential problems with using single-band dust
luminosities to estimate star formation rates in galaxies, due to the
varying influence of the ISM on the total dust luminosity. It may be
possible to mitigate these effects by making reasonable assumptions
about the SED, or by sampling wavelengths near the
temperature-dependent peak of the thermal dust emission. These are
important considerations given, for example, the very different
sensitivity at 450 and 850\,$\mu$m of SCUBA-2, and since the
850\,$\mu$m channel samples the 250\,$\mu$m rest-frame emission of
galaxies at $z \approx 2.4$ and 160\,$\mu$m rest-frame emission at $z
\gtsim 5.3$. Deriving monochromatic luminosities in this way may be
less susceptible to the effects of assuming an inappropriate dust SED
than using the integrated dust luminosity as a star formation rate
indicator.

\section{Conclusions}
\label{sec:conclusions}

We have used a 250\,$\mu$m-selected sample of 10,387 low-redshift
($z_{\mathrm{spec}} < 0.50$) galaxies from the \hatlas\ survey, with
isophotal major axes $< 10$ arc sec and spectroscopic redshifts, plus
aperture photometry at 1.4\,GHz based on data from the FIRST survey to
probe the far-infrared--radio correlation (FIRC). In order to
representatively probe the monochromatic FIRC, rather than focussing
only on the small sub-set of sources detected at 1.4\,GHz, we measure
aperture flux densities directly from the FIRST images for every
galaxy. This enables us to determine the FIRC by median stacking the
flux densities for galaxies in our sample as a function of parameters
of interest, as well as considering the individual galaxies with
formal detections.

We find that the monochromatic $q_{250}$ that we determine for
individual galaxies (i.e. those that are detected at $> 5\sigma$ in
both the {\it Herschel} and FIRST observations) are consistent with
expectations based on previous studies
\citep[e.g.][]{ivison10b,jarvis10}. In contrast, the stacked $q_{250}$
for our whole 250\,$\mu$m selected sample are offset to higher values
than those found in \citet{jarvis10}, highlighting the importance of
stacking techniques applied to large samples of sources. Though it is
possible that some fraction of our sample could contain low-level AGN,
the median stacking that we use, coupled with the fact that our
results are offset to higher $q_\lambda$ \citep[i.e. {\em lower}
  1.4\,GHz luminosities than the detections in our sample, or that
  of][]{jarvis10}, as well as our simulations (which suggest that the
influence of residual AGN contamination -- if it is present -- is
likely to be small), all offer encouragement in this regard.

Using the \hatlas\ catalogue and our FIRST flux densities we find no
evidence for redshift evolution of the FIRC as probed by the
$k$-corrected monochromatic 250\,$\mu$m luminosity density. This lack
of obvious evolution is in agreement with several previous studies
\citep[e.g.][]{boyle07,garn09a,garn09b,jarvis10,ivison10a,bourne11},
though we show this with a sample size unprecedented at these
redshifts for the first time.

In order to probe the temperature dependence of the monochromatic FIRC
we began by determining stacked FIRST luminosities in temperature
bins, revealing that the 1.4\,GHz luminosity increases as a function
of the effective dust temperature. This represents a radio continuum
version of the far-infrared Luminosity-Temperature (``L-T'') relations
discussed in several previous studies
\citep[e.g.][]{chapman03,hwang10,smith13,symeonidis13}. This cannot be
simply attributed to redshift or selection effects for two main
reasons. Firstly, unlike the 1.4\,GHz luminosity, we find that the
effective dust temperature of the stacked $k$-corrected PACS\slash
SPIRE SED in bins of redshift is constant. Secondly, our temperature
estimates \citep[along with the results of][]{symeonidis11} suggest
that our 250\,$\mu$m-selection includes galaxies with the vast
majority of dust temperatures at these redshifts (though we note that
we may miss galaxies with the very hottest temperatures $> 50$K, or
non-AGN radio-selected IR sources with low $q_\lambda$; it is unclear
how many of these exist at $z < 0.5$).

We use our stacked PACS\slash SPIRE\slash FIRST flux densities to show
that the monochromatic FIRCs, $q_\lambda$, show varying temperature
dependence. We find that $q_\lambda$ is roughly constant when sampling
near the peak of the dust SED (i.e. at 100 and 160\,$\mu$m), and that
it shows progressively stronger inverse correlation with temperature
as we move to the SPIRE observations at 250, 350 and 500\,$\mu$m. We
suggest that monochromatic far-IR data may be reliably used as star
formation rate indicators in particular situations, such as when the
observations sample wavelengths around 100\,$\mu$m in the rest-frame.
At these wavelengths, the fact that our results show a temperature
independent FIRC suggests that the far-IR SED is dominated by dust
heated in the most recent burst of star formation \citep[i.e. the
  stellar birth cloud component in the two model of][]{charlot00}
whatever the effective temperature of the far-IR SED.

The far-IR temperature\slash colour dependence of the FIRC is likely
to be of critical importance for future investigations, given the
impending explosion of Square Kilometre Array pathfinder and precursor
radio continuum surveys from ASKAP \citep{norris11}, LOFAR
\citep{rottgering11}, MeerKAT \citep{jarvis12}, and the Jansky Very
Large Array (JVLA). Radio observations will be crucial in the future,
since they will not only be sufficiently sensitive to detect the
entire $z < 0.5$ star-forming galaxy population, they may also provide
a more reliable tracer of a galaxy's star formation rate than
observations sampling long far-IR wavelengths ($> 250$\,$\mu$m). The
results in this paper and the data from these facilities are likely to
be critical for future studies of star-forming galaxies.

\section*{Acknowledgments}

The authors would like to sincerely thank the anonymous reviewer for
their thoughtful report which improved this paper. DJBS, MJJ, MV, NM
and MP also wish to thank the National Research Foundation of South
Africa for financial assistance. The authors would like to thank Chris
Hayward, Dominic Benford, Rob Ivison, Michal Micha{\l}owski and Paul
van der Werf for useful discussions. NB acknowledges funding from the
EC FP7 SPACE project ASTRODEEP (Ref. no. 312725).  EI acknowledges
funding from CONICYT/FONDECYT postdoctoral project
N$^\circ$:3130504. The {\it Herschel}-ATLAS is a project with {\it
  Herschel}, which is an ESA space observatory with science
instruments provided by European-led Principal Investigator consortia
and important with participation from NASA. The \hatlas\ website is
{\tt http:\slash \slash www.h-atlas.org\slash}. GAMA is a joint
European-Australasian project based around a spectroscopic campaign
using the Anglo-Australian Telescope. The GAMA input catalogue is
based on data taken from the Sloan Digital Sky Survey and the UKIRT
Infrared Deep Sky Survey. GAMA is funded by the STFC (UK), the ARC
(Australia), the AAO, and the participating institutions. The GAMA
website is http:\slash \slash www.gama-survey.org\slash. This work
used data from the SDSS DR7. Funding for the SDSS and SDSS-II has been
provided by the Alfred P. Sloan Foundation, the Participating
Institutions, The National Science Foundation, the U.S. Department of
Energy, the National Aeronautics and Space Administration, the
Japanese Monbukagakusho, the Max Planck Society and the Higher
Education Funding Council for England.

\bibliographystyle{mn2e}\bibliography{parsley_refs}

\appendix

\section{More SED fits and $k$-corrections} 
\label{sec:kcor}

\subsection{Integrated dust luminosities}

Since isothermal models do not include any mid-IR contribution to the
total dust luminosity from warmer dust components, such as hot very
small grains \citep[e.g.][]{yang07}, we have used three alternative
methods to try and quantify the likely impact that our choice of SED
will have on our derived dust luminosities\slash $k$-corrections. The
additional methods we use to calculate \ldust\ are described below:

\begin{itemize}
  \item As discussed in section \ref{subsubsec:mono_estimates}, we fit
    our PACS and SPIRE data using the panchromatic SED templates from
    \citet[][hereafter SK07]{sk07}. Though \citet{smith12} found that
    these templates did not reproduce the optical\slash near-IR
    properties of \hatlas\ galaxies, the SK07 templates are the only
    set of available multi-component templates that include models
    cold enough to describe the far-IR SEDs of sources with $\teff <
    15$\,K that we find in \hatlas. We derive the best fits based on
    these models using the PACS\slash SPIRE data alone.
  \item We also use results based on \magphys\ \citep[][hereafter
    DC08]{dacunha08}; a panchromatic SED fitting code which assumes
    consistency between the energy absorbed by dust \citep[using a
      two-component obscuration model from][]{charlot00}, and the
    energy reradiated in the far-infrared. \magphys\ produces best-fit
    and median-likelihood estimates of dust luminosity in the same way
    as our isothermal fitting, and the application of \magphys\ to the
    \hatlas\ data set is described in great detail in
    \citet[][hereafter S12]{smith12}. We note that the coldest dust
    temperature component included in the DC08 dust library is 15\,K,
    meaning that \magphys\ is unable to accurately reproduce the dust
    SEDs of the minority of sources colder than this using the
    standard priors. We also note that the temperatures of the two
    dust SED components included in \magphys\ do not, in general, map
    onto the effective temperatures that we derive for our sample
    based on the isothermal model.
  \item We also derived estimates of \ldust\ using the DC08 far-IR SED
    template library without using the optical\slash near-IR data,
    i.e. without using the energy balance criterion imposed, deriving
    \ldust\ estimates in the same way as for the SK07 models,
    above. We refer to these values as the DC08 results in what
    follows.
\end{itemize}

We find that there are temperature-dependent offsets between the
integrated dust luminosities derived using each of these methods;
``correcting'' the isothermal values to total integrated dust
luminosities is highly temperature\slash model dependent. For example,
we are unable to assume a simple correction factor to convert the
isothermal dust luminosity to agree with the \magphys\ estimates, as
it is a function of temperature, with significant scatter.

In this investigation we would also like to probe the temperature
dependence of the integrated FIRC \citep[i.e. to update and build on
  the study by][in light of the newly-available PACS data, $10\times$
  larger areal coverage and additional spectroscopic
  redshifts]{jarvis10}. The only previous investigation of this
dependence \citep{ivison10a} found variations on the scale of $\sim
0.1$\,dex over the temperature range probed in \hatlas, but the
uncertainties on even the \magphys\ dust luminosities (due to the
variable contribution from the mid-infrared) are larger than
this. Furthermore, the differences between the dust luminosities
derived using the different SED fits discussed above are compounded by
the absence of sensitive mid-infrared data available for our SED
fitting at the time of writing \citep[as highlighted by][and discussed
  in section \ref{sec:ldust_est}]{smith12}. As a result, we are unable
to address the temperature dependence of the integrated FIRC at this
time, and so leave this topic for a future study.

\subsection{Uncertainties on $k$-corrections}

\begin{figure}
  \centering
  \includegraphics[width=0.92\columnwidth]{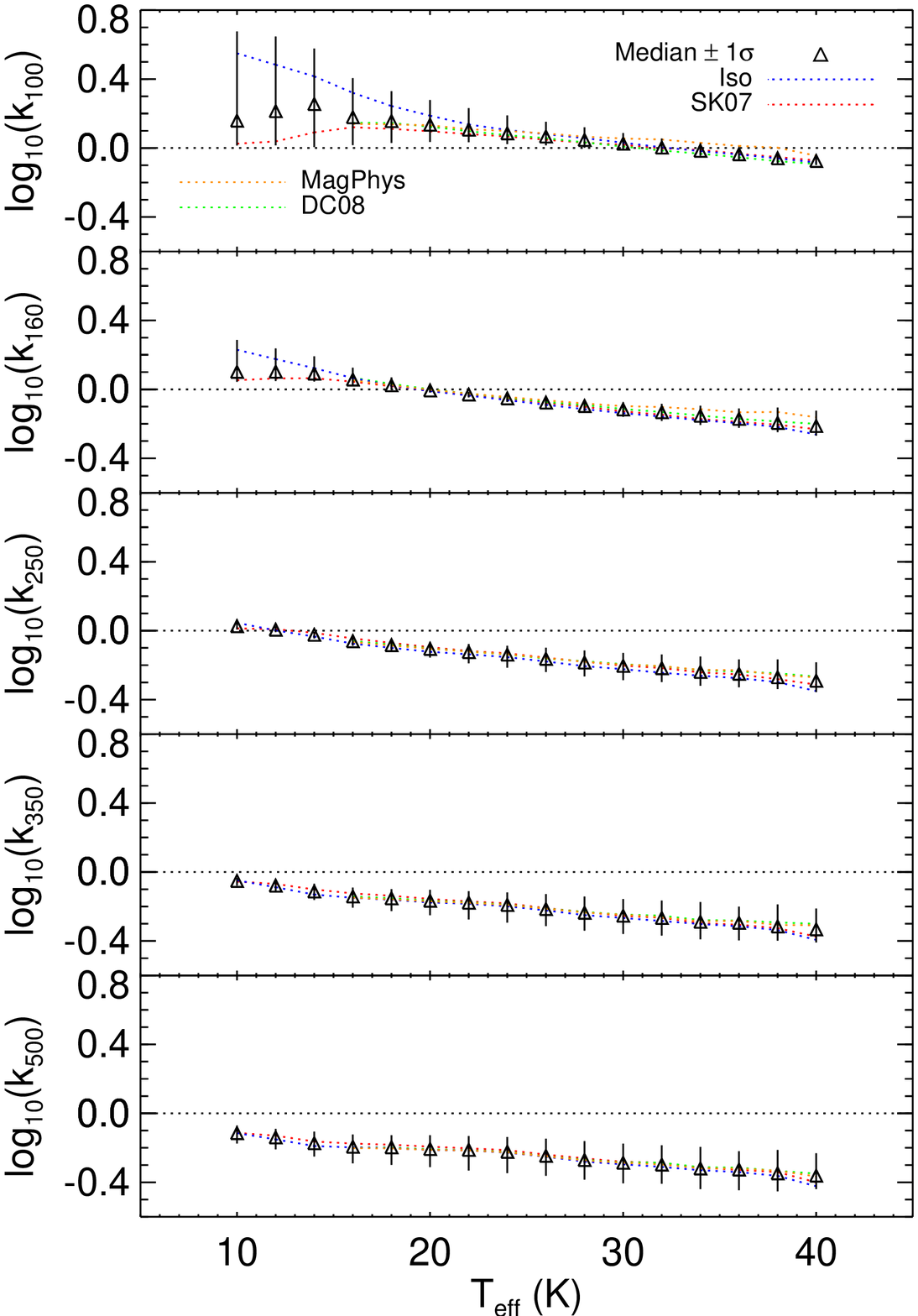}
  \caption{The median $k$-corrections ($k_\lambda$; triangles) at each
    of the PACS\slash SPIRE wavelengths as a function of temperature,
    conservatively averaged over the sets of SED fits, and over
    redshift. The uncertainty on $k_\lambda$ as a function of
    \teff\ is shown by the error bars, with median values of 25.3,
    11.6, 14.3, 18.0 and 21.4\,per cent in each of the PACS\slash
    SPIRE bands in increasing order of wavelength. The largest
    uncertainties rise to a factor of $\sim 3$ on $k_{100}$ at $\teff
    \approx 10$\,K. The coloured dotted lines indicate the median
    $k_\lambda$ using each individual set of templates, with the
    colour-coding indicated in the legends. }
  \label{fig:sigma_k}
\end{figure}

We generate conservative estimates of the uncertainties on the
monochromatic $k$-corrections by first calculating $k_\lambda$ for
every source using each of the four sets of template SEDs (isothermal,
SK07, \magphys\ and DC08). We then bin our sample by isothermal
temperature, and calculate the standard deviation, $\sigma
\left(k_\lambda\right)$, across the bin occupants. In calculating
$\sigma \left(k_\lambda\right)$ for a particular temperature bin, we
include the best-fit $k_\lambda$ to each galaxy derived using each of
the aforementioned SED libraries (providing that they have reduced
$\chi^2 < 2$). 

We suggest that the resulting estimates of $\sigma \left(k_\lambda
\right)$ are likely to be conservative (i.e. over-estimated) for two
reasons. Firstly, the range of SEDs in the isothermal, SK07 and DC08
libraries is probably larger than the range of SEDs of star-forming
galaxies in \hatlas, particularly at $\lambda_{\mathrm{obs}} <
100\,\mu$m where we use the templates to extrapolate beyond the
observational data, and where we know that the isothermal models
underestimate the true SED. Secondly, we calculate global values as a
function of temperature alone, rather than calculating
$\sigma(k_\lambda)$ as a function of redshift (i.e. we do not
discriminate between our differing ability to determine temperatures
for e.g. $\teff = 20$\,K galaxies at $z = 0.5$ compared to galaxies
with the same temperature at $z = 0.05$). Finally, we note that in
deriving $\sigma \left(k_\lambda \right)$, we include four sets of
templates above $\teff = 15$\,K, while at colder temperatures we
include only the isothermal and SK07 templates in the averaging, since
the standard \magphys\ libraries used for the other two sets of fits
do not include dust SED components colder than 15\,K.

The median $\sigma \left(k_\lambda\right)$, shown in figure
\ref{fig:sigma_k}, range from $\sim 14$ per cent at 250\,$\mu$m (where
the dust SED is best-sampled) to $\sim 25$ per cent at 100$\mu$m (at
the edge of our far-IR wavelength coverage, and where the different
SEDs show most variation), though the uncertainty on $k_{100}$ at the
coldest temperatures is rather larger. We propagate the uncertainties
shown in figure \ref{fig:sigma_k} onto our dust luminosity estimates
by adding them in quadrature with the uncertainties on the flux
densities in the \hatlas\ catalogue; these are then propagated through
onto the individual $q_\lambda$.

\bsp

\label{lastpage}

\end{document}